\documentstyle[12pt]{article}

\def\fun#1#2{\lower3.6pt\vbox{\baselineskip0pt\lineskip.9pt
        \ialign{$\mathsurround=0pt#1\hfill##\hfil$\crcr#2\crcr\sim\crcr}}}

\renewcommand\({\left(}
\renewcommand\){\right)}

\newcommand\eq[1]{Eq.~(\ref{#1})}
\newcommand\eqs[2]{Eqs.~(\ref{#1}) and (\ref{#2})}
\newcommand\eqss[3]{Eqs.~(\ref{#1}), (\ref{#2}) and (\ref{#3})}

\newcommand\eqst[2]{Eqs.~(\ref{#1})--(\ref{#2})}

\newcommand\ee{\end{equation}}
\newcommand\be{\begin{equation}}
\newcommand\eea{\end{eqnarray}}
\newcommand\bea{\begin{eqnarray}}

\newcommand\TeV{\,\mbox{TeV}}
\newcommand\GeV{\,\mbox{GeV}}
\newcommand\MeV{\,\mbox{MeV}}
\newcommand\keV{\,\mbox{keV}}

\newcommand\mpl{M_{\rm P}}

\newcommand\lsim{\mathrel{\rlap{\lower4pt\hbox{\hskip1pt$\sim$}}
    \raise1pt\hbox{$<$}}}
\newcommand\gsim{\mathrel{\rlap{\lower4pt\hbox{\hskip1pt$\sim$}}
    \raise1pt\hbox{$>$}}}

\newcommand\diff{\mbox d}

\def\dslash{\not{\hbox{\kern-2pt $\partial$}}}
\def\Dslash{\not{\hbox{\kern-4pt $D$}}}
\def\Oslash{\not{\hbox{\kern-4pt $O$}}}
\def\Qslash{\not{\hbox{\kern-4pt $Q$}}}
\def\pslash{\not{\hbox{\kern-2.3pt $p$}}}
\def\kslash{\not{\hbox{\kern-2.3pt $k$}}}
\def\qslash{\not{\hbox{\kern-2.3pt $q$}}}

 \newtoks\slashfraction
 \slashfraction={.13}
 \def\slash#1{\setbox0\hbox{$ #1 $}
 \setbox0\hbox to \the\slashfraction\wd0{\hss \box0}/\box0 }
 
\def\ee{\end{equation}}
\def\be{\begin{equation}}

\newcommand\sub[1]{_{\rm #1}}

\newcommand\mgrav{m_{3/2}(t)}
\newcommand\mgravsq{m_{3/2}(t)}
\newcommand\mgravvac{m_{3/2}}

\begin{document}

\begin{flushright}
LANCS-TH/9922  
\\hep-ph/9911257\\
(November 1999)
\end{flushright}
\begin{center}
{\Large \bf The gravitino abundance in supersymmetric \\
`new' inflation models}

\vspace{.3in}
{\large\bf  David H.~Lyth}

\vspace{.4 cm}
{\em Department of Physics,\\
Lancaster University,\\
Lancaster LA1 4YB.~~~U.~K.}

\vspace{.4cm}
\end{center}

\begin{abstract}
We consider  the abundance of gravitinos
created from the vacuum fluctuation, in a class of `new' inflation
models for which global supersymmetry is a good approximation.
Immediately after inflation, gravitinos are produced, with  number density
 determined by  equations 
recently presented by Kallosh et. al.
(hep-th/9907124) and Giudice et. al.
(hep-ph/9907510). 
Unless reheating intervenes, creation may continue, maintaining 
about the same number density,
until the Hubble parameter falls below the  gravitino mass.
In any case, the abundance of gravitinos created from the vacuum
fluctuation
exceeds the abundance from thermal collisions in a significant
regime of parameter space, leading to
tighter cosmological constraints.
\end{abstract}

\paragraph{Introduction} 
Gravitinos are created in the early Universe with a 
cosmologically significant abundance. They are certainly created by 
thermal collisions after reheating \cite{subir}, and some time ago
 \cite{lrs}
it was pointed out
that they may also be created non-thermally, starting  from 
the vacuum fluctuation that exists  well before horizon exit during
 inflation.
 It was conjectured that the initial  gravitino number density from
 this mechanism 
 would be of order $\mgrav^3$, where  the effective  gravitino
mass $\mgrav$ is at most of order the Hubble parameter.
 In that case, gravitino 
creation from the vacuum is insignificant
compared with creation from thermal collisions.

Recently, the mode function equations determining the gravitino
abundance have been worked out \cite{kklv,grt}, for models
in which  a single real scalar field dominates the density
and pressure of the Universe.
On the basis of these equations, their authors have pointed out
that  gravitinos are  created just after inflation with number
density of order $M^3$, where $M$ is the mass of the inflaton.
In many models of inflation, this already makes gravitino creation
from the vacuum much more important than creation from thermal
collisions. Furthermore, creation may continue, maintaining about
the same number density, until  the Hubble parameter falls below the
true 
gravitino mass $\mgravvac$ 
\cite{p99new2}, making gravitino creation from the vacuum
even more significant. 

In this note, we consider a 
specific class \cite{kmy} of supersymmetric `new' inflation  models.
We estimate the abundance of gravitinos for both the
minimal case, where creation ends soon after inflation, and the maximal case
where it continues until the Hubble parameter falls below the gravitino mass.

\paragraph{The model}
The models that we consider invoke a
tree-level supergravity theory, containing no physical fields except
the gravitino, and a complex scalar field $\phi_1$
with the minimal kinetic term. (The degrees of freedom
 corresponding to
the spin $1/2$ partner of the scalar field are eaten by the helicity
$1/2$ components of the gravitino field.)
There is a  holomorphic superpotential
$W(\phi_1)$, leading to the potential 
\bea
V &=& F^2 -3\mpl^2 |\mgravsq|^2 \label{sugra}\\
F^2&\equiv&  e^{|\phi_1|^2/\mpl^2}
\left| \frac{\diff W}{\diff \phi_1}
+\mpl^{-2}\phi_1^* W \right|^2  \\ 
\mgrav&\equiv & e^{\frac12|\phi_1|^2/\mpl^2}W/\mpl^2
\label{gmass}
\,.
\eea
In the vacuum, $\mgrav$ is the gravitino mass, denoted $\mgravvac$
without an argument.

The  superpotential is  of the form \cite{kmy}
\be
W = V_0^\frac12 \phi_1 \( 1- \frac1{p+1} \(\sqrt 2 \phi_1/v\)^p 
\)
\,,
\ee
with $p\geq 3$. 
The real parameter $v$ is taken
to be small on the Planck scale, $v\ll \mpl$ where $\mpl =2.4
\times 10^{18}\GeV$. 

Because $v$ is real,  ${\rm Im}\;\phi_1$ is driven to 
zero, leaving the canonically-normalized inflaton field
$\phi=\sqrt 2 {\rm Re\;} \phi_1$.
Since $W$ has no constant term, the quadratic terms
in \eq{sugra} cancel. At $|\phi_1|\ll \mpl$, 
global supersymmetry is a good approximation except near the minimum of 
the potential.\footnote
{For $p\geq 5$ one also needs $\phi/v\gg (v/\mpl)^\frac4{p-4}$,
which is assumed.} This gives
\be
V\simeq F^2\simeq \left|\frac{\diff W}{\diff \phi_1}\right|^2
=V_0 \(1 - (\phi/v)^p \)^2
\,.
\label{v}
\ee

The field equation is
\be
\ddot \phi + 3H\dot\phi + V' =0
\label{feq}
\,
\ee
where $H$ is the Hubble parameter, related to the energy density by
$3\mpl^2 H^2=\rho$. The density and pressure are 
\bea
\rho &=& V +\frac12\dot\phi^2 \label{rho}\\
P &=& -V +\frac12\dot\phi^2 \label{P}
\,,
\eea
with $\dot \rho=-3H(\rho+P)$.

Inflation occurs in the regime $\phi\ll v$, while $\phi$ is rolling away 
from the origin. The vacuum expectation value (vev) of $\phi$ is 
precisely $v$ in this approximation, and $V=0$ corresponding to
unbroken supersymmetry. The
mass of $\phi$ in the vacuum is
\be
M\sim \frac{V_0^\frac12}{v } \sim H_* \(\frac {\mpl}{v} \) \gg H_*
\label{mofv1} 
\,,
\ee
where $H_*$ is the Hubble parameter at the end of inflation.

Before proceeding, we note that this class of inflation  models
is reasonably well-motivated.
The form of the superpotential may be motivated by invoking a
$Z_p$ symmetry ($R$-symmetry). Such a symmetry 
allows additional terms only 
of order $(\phi_1/v)^{1+np}$ ($n\geq 2$). 
The assumption of a practically minimal kinetic function (K\"ahler 
potential) $K=|\phi_1|^2$ is not completely unreasonable, since
of the expected higher-order terms $\sim \mpl^{2-2n} |\phi_1|^{2n}$
only the first need be suppressed \cite{kmy}.
The main limitation on the model is the requirement that the neglected 
fields all have vevs much less than $\mpl$; indeed, just one field
of order
$\mpl$, with the minimal kinetic term,
will make the potential too steep for inflation
\cite{treview}, and there is no reason why non-minimal terms 
or additional fields should flatten the potential.

Near the minimum of  the potential, corresponding to the vacuum
of the one-field theory under consideration, we have to use the supergravity
expression \eq{sugra}. This gives negative  vacuum energy,
\bea
V\sub{vac}&=& F\sub{vac}^2 - 3\mpl^2 \mgravvac^2 \label{vvac} \\
3\mpl^2\mgravvac^2 &\simeq& -3V_0 (v/\mpl)^2 \\
F\sub{vac}^2  &\sim& V_0 (v/\mpl)^4 
\,.
\eea

In the true  vacuum, the potential
 (practically) vanishes. This means that the true vacuum values
of  $F$ and/or the gravitino mass  must be generated by
some other sector of the Lagrangian, than the one used for the model
of inflation. One hypothesis \cite{kmy} is that the model gives 
 the true gravitino  mass, with the additional sector
generating only the true  vacuum value of $F$
which as usual we denote by  $M_S^2$.
 For definiteness we adopt this hypothesis,
which actually seems the most natural in view of the 
requirement  that there be no Planck-scale vevs, at least during inflation.
(To implement supersymmetry breaking without Planck-scale vevs, 
one might invoke a gauge-mediated mechanism or 
a Fayet-Iliopoulos term, neither of which would significantly
affect the gravitino mass.)

We shall make estimates for the cases $p=3$, $p=4$ and 
$p\gg 2$.
A relation between $V_0$ and $v$ 
is provided by the COBE measurement of the cosmic microwave background
anisotropy \cite{treview},
\be
X(p)
=
\(\frac {\mpl}{v} \)^\frac p{p-2} \(\frac {V_0^{\frac12}}{\mpl^2} \)
\label{cobe}
\,,
\ee
where
\be
X\equiv 
5.3\times 10^{-4}p^{-\frac1{p-2}} [N(p-2)]^{-\frac{p-1}{p-2}}
\,.
\ee
Here $N$ is the number of $e$-folds of slow-roll inflation after
cosmological scales leave the horizon. We take $N=50$, leading to
the following estimates.
\bea
&p=3:\ &\frac v\mpl \simeq 10^2\( \frac{V_0^\frac14}{\mpl} \) ^\frac23
\\
&p=4:\ &\frac v\mpl \simeq 10^3 \( \frac{V_0^\frac14}{\mpl} \) 
\\
&p\gg 2:\ &\frac v\mpl \simeq 10^5 p \( \frac{V_0^\frac14}{\mpl} \) ^2
\,.
\eea
This determines the potential in terms of $\mgravvac$,
as follows.
\bea
&p=3:\ &V_0^\frac14 \sim \(\frac{\mgravvac}{100\GeV} \)^{3/8}\times
10^{11}\GeV 
\label{mt1}\\
&p=4:\ &V_0^\frac14 \sim \(\frac{\mgravvac}{100\GeV} \)^{1/3}\times
10^{12}\GeV 
\label{mt2}\\
&p\gg 2:\ &V_0^\frac14 \sim \(\frac{\mgravvac}{100\GeV} \)^{1/4}
p^{-1/4}\times
10^{13}\GeV \label{mt3}
\,.
\eea
If $\mgravvac$ is the true gravitino mass, it presumably lies roughly
in the range $1\keV$ to $100\GeV$.

\paragraph{The helicity $1/2$ gravitino mass}
The model contains only a single chiral superfield, with minimal kinetic term,
which can be taken to be real.
The gravitino field therefore
 obeys \cite{kklv,grt} the Rarita-Schwinger equation,
with a time-dependent mass given by \eq{gmass}, and constraints
to eliminate unphysical degrees of freedom. The equation and the 
constraints have to be evaluated in the curved spacetime corresponding 
to the expanding Universe. This gives separate mode function equations for
the helicity $1/2$ and helicity 
$3/2$ states, as seen by a comoving observer.

The helicity $3/2$ mode function satisfies the same equation 
as a spin $1/2$
particle with mass $\mgrav$ \cite{mm,kklv,grt,l}.
There is practically no creation of helicity $3/2$ gravitinos
 from the vacuum in the present model, corresponding to the conformal 
invariance of the massless Dirac equation.

The helicity $1/2$ mode function,
satisfies the same equation as a spin $1/2$ particle with mass
\cite{kklv,grt}\footnote
{This is the expression given in \cite{kklv}.
The result of \cite{grt} leads to an identical expression \cite{nont},
except that the last term of \eq{mtil} is $-\mu$. 
This discrepancy 
does not affect order of magnitude calculations.}
\be
\tilde m(t)= \mgrav -\frac32 m (1+A_1) -\frac32 H A_2 + \mu 
\,,
\label{mtil}
\ee
where
\bea
A_1 &\equiv& \frac{P-3\mpl^2 \mgravsq}{\rho+3\mpl^2\mgravsq}\equiv \cos \chi
\label{A1} \\
A_2&\equiv &\frac23\frac{3\mpl\dot \mgrav}{\rho+3\mpl^2\mgravsq} 
\equiv\sin\chi \label{A2}\\
A&\equiv& A_1 + i A_2 = e^{i\chi} \\
\mu&\equiv & -\frac12 \dot\chi
\,.
\eea
Using \eqss{sugra}{rho}{P}, we can write \eq{A1} as
\be
A_1 = \frac{\frac12\dot\phi^2-F^2}{\frac12\dot\phi^2+F^2}
\label{A1ofF}
\,.
\ee

As already noted, the helicity $1/2$ components of the 
gravitino field 
have, in this model, the same dynamics as 
a spin $1/2$ field with effective mass $\tilde m$.
They are produced with
momentum $k/a$ if 
there is appreciable violation of a weak adiabaticity condition
\cite{nont}
\be
|\overline{(a\tilde m)'}| \ll \omega^2\equiv k^2 + (a\tilde m)^2
\,,
\ee
where the prime denotes differentiation with respect to conformal time,
$\diff/\diff\eta=a \diff/\diff t$, 
and the average is over a conformal time interval
$\omega^{-1}$. 
(As usual, $a$ is the scale factor of the Universe, $H=\dot a/a$.)
 In practice
$k\sub{max}$, the biggest $k$ for which significant creation occurs,
is simply the biggest value achieved by $a\tilde
m$, within the regime
where $\tilde m$ varies non-adiabatically ($|\dot {\tilde m}
|\gsim \tilde m^2$).

To estimate $k\sub{max}$ in this model, we need to 
 follow the evolution of  $\tilde m$. During inflation, it is 
 slowly varying, with 
$\tilde m\simeq \mgrav \ll H_*$.
After inflation, $\phi$ oscillates 
about its vev. Except  during the first few Hubble
times, the amplitude $\phi_0$ is much less than $v$, 
and $\phi-v\simeq \phi_0 \sin
 Mt$. As long as global supersymmetry
is a good approximation, this gives
\bea
F^2 &\simeq& \frac12 M^2\phi_0^2 \sin^2 Mt \label{F}\\
\frac12\dot\phi^2 &\simeq& \frac12 M^2 \phi_0^2 \cos^2 Mt
\,.
\eea
{}From \eq{A1ofF}, this gives $A_1=\cos 2Mt$ and therefore
\be
\tilde m(t) \simeq \mu(t) \simeq -M \,.
\ee
(The overall sign of  $\tilde m$ is determined by \eqs{gmass}{A2}, but it is not
physically significant.)
The mass $M$ of the inflaton is also the mass of the  inflatino.
	In the model under consideration the latter, in turn,
becomes the goldstino in the limit of global supersymmetry.\footnote
{The   goldstino is the fermionic component of the chiral superfield
responsible for spontaneous global supersymmetry breaking.
In the early Universe, when the time-derivative of the scalar
component, not just the $F$ component, contributes to 
 supersymmetry breaking,  the goldstino mass does not  vanish \cite{grt2}
 (see also
\cite{mp}).}
This is the  physical reason why $\tilde m$ rises sharply in magnitude
 just after inflation. The sharp rise creates gravitinos,
with $k\sub{max}\sim a_* M$, where a star denotes
the end of inflation.

 Taking this model literally, the negative vacuum
energy means that the Hubble parameter falls to zero in the early
Universe,
\bea
3\mpl^2H^2 &=&\rho(t) \simeq M^2\phi_0^2(t) + V\sub{vac}\\
&\simeq& M^2 \phi_0^2(t) - 3\mpl^2 \mgravvac^2 
\,.
\label{rhomod}
\eea
When $H$ vanishes, the Universe attains its maximum size. After that, it
recollapses.
Global supersymmetry
is a good approximation for $F$ except near $F=0$,
 because its amplitude $M\phi_0$ never falls below
$\sqrt 3\mpl\mgravvac \gg F\sub{vac} $. As a result, $\tilde m$ maintains
the constant value $M$, and gravitino production occurs only
at the end of inflation. Taking the model literally, we therefore
have 
\be
k\sub{max} \sim a_* M
\label{minimal}
\,.
\ee
where   the star denotes the end of inflation.

In reality, the energy density is not given by \eq{rhomod},
but by 
\be
3\mpl^2H^2 =\rho(t) \simeq M^2\phi_0^2(t) 
\,,
\ee
with the negative inflaton contribution to the vacuum energy canceled
by whatever field is responsible for supersymmetry breaking in the true
vacuum. It is argued elsewhere \cite{p99new2} that as a result,
gravitino creation will  continue unless reheating intervenes,
with every-increasing $k\sub{max}\sim aM$, until
 the Hubble parameter falls below the true gravitino mass.
In  the present model this epoch corresponds to
\bea
\(\frac{a_*}{a} \)^\frac34 \sim
\frac{\rho^\frac14}{V^\frac14_0}
 &\sim& \(\frac{\mgravvac}{H_*} \)^\frac12 \\
&\sim&\( \frac v\mpl \) ^\frac12 
\label{epoch4}
\,,
\eea
leading to 
\be
k\sub{max} \sim a_* M (\mpl/v)^\frac23
\label{maximal}
\,.
\ee

\paragraph{The abundance of gravitinos created from the vacuum}
Since we are dealing with fermions, the occupation number 
of each helicity state
is $\leq 1$. It is expected to be of order 1 at $k=k\sub{max}$, 
giving number density \cite{lrs}
\bea
n &\simeq& \frac2{4\pi^2} a^{-3} \int^{k\sub{max}}_0 k^2 dk\\
&\simeq &10^{-2} a^{-3} k\sub{max}^3 
\,,
\eea
The relative abundance at 
nucleosynthesis is \cite{lrs}
\be
\frac ns \simeq 10^{-2} \(\frac{k\sub{max}}{a_*}\)^3
\frac {\gamma T\sub R}{V_0}
\,,
\ee
where $s$ is the entropy density at nucleosynthesis,
and $\gamma^{-1}$ is the increase in entropy per comoving volume
(if any), between reheating at
temperature $T\sub R< V_0^\frac14$ and nucleosynthesis. 

In the present model, a  minimal estimate for the gravitino 
abundance is provided by \eq{minimal}, whereas a maximal 
one is provided by \eq{maximal}. Let us suppose first that the
 maximal estimate \eq{maximal} is correct.
Depending on the value of $p$, we find
\bea
&p=3:&\ \frac ns \sim 10^{-10} \(\frac \mpl {V_0^\frac14} \) ^\frac43
\frac{\gamma T\sub R}{\mpl}
\label{34}
\\
&p=4:&\ \frac ns \sim 10^{-15} \(\frac \mpl {V_0^\frac14} \) ^3
\frac{\gamma T\sub R}{\mpl}
\label{35}
\\
&p\gg 2:&\ \frac ns \sim 10^{-27} p^{-5} \(\frac \mpl {V_0^\frac14} \) ^8
\frac{\gamma T\sub R}{\mpl}
\label{36}
\,.
\eea

The cosmological significance \cite{subir} of the gravitino depends on its 
true mass $\mgravvac$.
A gravitino with mass more than a few  $\TeV$ has no effect
because it decays well before nucleosynthesis, but such a big mass is 
regarded as unlikely. 

A gravitino with mass in the range $100\MeV\lsim \mgravvac\lsim \TeV$
decays around or after 
nucleosynthesis, but before the present.
This range includes the value $\mgravvac\sim 100\GeV$ to $1\TeV$,
expected in gravity-mediated models of supersymmetry breaking.
Observation then requires \cite{subir}
\be
n/s\lsim 10^{-13}
\,. 
\label{nsyn}
\ee
(To be precise, the 
upper bound depends on the mass and is in the range
$10^{-12}$ to $10^{-15}$.) 
The abundance of gravitinos from thermal 
collisions is $n/s\sim 10^{-13} (\gamma T\sub R/10^9\GeV)$, leading to
the bound $\gamma T\sub R\gsim
10^9\GeV$. Using instead \eqst{mt1}{mt3} and \eqst{34}{36}, we find
the following results.
\bea
&p=3:&\ \gamma T\sub R\lsim 10^6\GeV
\\
&p=4:&\ \gamma T\sub R\lsim 100\GeV
\\
&p\gg 2:&\ \gamma T\sub R\lsim 10^{-25}\GeV
\,.
\eea

A gravitino with mass $\mgravvac\lsim 100\MeV$
survives to the present,
and is a dark matter candidate. This
includes the range predicted by gauge-mediated
models of supersymmetry breaking, which is $1\keV\lsim \mgravvac\lsim
100\GeV$ with the upper decades disfavoured.
The present density is
\be
\Omega_{3/2} \simeq 10^5 \( \frac{\mgravvac}{100\keV } \)
\frac ns 
\ee
If creation from thermal collisions dominates, 
then very roughly \cite{subir}
\be
\Omega_{3/2} \sim\(\frac{100\keV}{\mgravvac} \) \(\frac
{T\sub R}{10^4\GeV} \) 
\,.
\ee
Using instead
\eqst{mt1}{mt3} and \eqst{34}{36}, we find the following results
for $\mgravvac=100\keV$;
\bea
&p=3:&\ \gamma T\sub R\lsim 10^{11}\GeV
\\
&p=4:&\ \gamma T\sub R\lsim 10^4\GeV
\\
&p\gg 2:&\ \gamma T\sub R\lsim 10^{-17}\GeV
\,.
\eea

These results are quite dramatic. The  bound
on the reheat temperature is generally stronger than the one coming from
thermal collisions, indicating that gravitino production from the vacuum
is dominant. For $p\gg 2$, the abundance is so big that an epoch
of thermal inflation \cite{thermal} would be needed.

Finally, suppose that  the minimal estimate \eq{minimal} is correct.
Then we have
\bea
&p=3:&\ \frac ns \sim 10^{-8} 
\frac{\gamma T\sub R}{\mpl}
\\
&p=4:&\ \frac ns \sim 10^{-11} \(\frac \mpl {V_0^\frac14} \) 
\frac{\gamma T\sub R}{\mpl}
\\
&p\gg 2:&\ \frac ns \sim 10^{-15} p^{-3} \(\frac \mpl {V_0^\frac14} \) ^4
\frac{\gamma T\sub R}{\mpl}
\,.
\eea
In this minimal case, there is no regime of parameter space where
gravitinos created from the vacuum are dominant, in the simplest
case that the model is responsible for the true gravitino mass.
In the more general case, they are again dominant in a significant
region of parameter space.

\paragraph{Conclusion}
In a specific class of 
inflation models, giving a potential $V\simeq V_0(1-(\phi/v)^p)$,
we have made minimal and maximal estimates, for  the abundance of 
gravitinos created from the vacuum fluctuation.
In the  maximal case, gravitinos from this mechanism may be so abundant
that an era of thermal inflation is  needed to dilute them.
In the minimal case, gravitinos from the vacuum fluctuation 
have no significant effect, if the gravitino mass given by the model
is the true mass.

\subsubsection*{Acknowledgements}
I am indebted to Toni Riotto and Andrei Linde for useful discussions.

\newcommand\pl[3]{Phys. Lett. #1 (19#3) #2}
\newcommand\np[3]{Nucl. Phys. #1 (19#3) #2}
\newcommand\pr[3]{Phys. Rep. #1 (19#3) #2}
\newcommand\prl[3]{Phys. Rev. Lett. #1 (19#3) #2}
\newcommand\prd[3]{Phys. Rev. D #1 (19#3) #2}
\newcommand\ptp[3]{Prog. Theor. Phys. #1 (19#3) #2}
\newcommand\rpp[3]{Rep. on Prog. in Phys. #1 (19#3) #2}
\newcommand\jhep[2]{JHEP #1 (19#2)}
\newcommand\grg[3]{Gen. Rel. Grav. #1 (19#3) #2}

\end{document}